\documentclass{jltp}
\usepackage{graphicx}
\newcommand{\p}[1]{(\ref{#1})}

\title{On a
Theory of Multi-gap Superfluidity \\Based on  Fermi--liquid Approach}

\author{A.I. Akhiezer,
A.A. Isayev,
S.V. Peletminsky and\\
A.A. Yatsenko}

\address{Kharkov Institute of Physics and Technology,
  Kharkov, 310108, Ukraine
}

\runninghead{A.I. Akhiezer \it{et al.}}{Theory of  Multi-gap
Superfluidity }

\begin {document}
\maketitle
\begin{abstract}
Superfluid-to-superfluid phase transitions in a Fermi-liquid leading to
the emergence of two--gap superfluid states have been studied.
  There are considered the
systems of fermions of one and two (nuclear matter) sorts.  The
self-consistent equations describing new two-gap superfluid states
have been obtained. These equations differ essentially from the equations of
the BCS theory. The solutions of the self-consistent equations
corresponding to one--gap and two--gap superfluid states have been
found.

\end{abstract}

\section{INTRODUCTION}
As a rule, superfluid states in a Fermi-liquid (FL) are considered as
arising as a result of a phase transition from the normal state. However,
phase transitions that might also take place in a superfluid FL can lead
to the emergence of new superfluid states. Thus, we are speaking of phase
transitions from one superfluid state to another. A new superfluid state can
be characterized by not one but a few order parameters. In this report we
consider the systems of fermions of one and two sorts. In the first case
the possibility of a phase transition to the state for which the spin of a
Cooper pair has no definite value but is equal with certain probability to
zero or unity is studied (singlet-triplet pairing of fermions). In the
second case a phase transition in superfluid nuclear matter to the state
corresponding to the superposition of states with singlet-triplet and
triplet-singlet pairing of nucleons (in spin and isotopic spaces) is
considered.

\section{SYSTEMS OF FERMIONS OF ONE SORT}
Our
analysis is based on the theory of a superfluid FL \cite{AKP}.
				       A Fermi
superfluid is described by  the normal $f_{\kappa_1\kappa_2}=\mbox{Tr}\,
\varrho
a^+_{\kappa_2}a_{\kappa_1}$ and anomalous  $g_{\kappa_1\kappa_2}=\mbox{Tr}\,
\varrho
a_{\kappa_2}a_{\kappa_1}$ distribution functions ($\varrho$ is the density
matrix of
the system,  $a^+_\kappa$ and $a_\kappa$ are the creation and annihilation
operators of a fermion in the state with momentum $\mathbf p$, spin
projection $\sigma$; for a nucleon in nuclear matter the state is
characterized also by the isotopic
spin projection $\tau$).  The energy of the system,
$E=E(f,g)$,  determines the fermion one-particle energy
 and the matrix
order parameter   \begin{equation}
\varepsilon_{\kappa_1\kappa_2}=\frac{\partial E}
{\partial f_{\kappa_2\kappa_1}},\quad
\Delta_{\kappa_1\kappa_2}=2\frac{\partial E}{\partial g_{\kappa_2\kappa_1}^+}
\label{1}\end{equation} The distribution functions $f$ and
$g$ are related with the
quantities $\varepsilon$ and $\Delta$ by the formulas
\cite{AIP}:  \begin{equation}
f=\frac{1}{2}\left(1-\frac{\xi}{E}\tanh\frac{Y_0E}{2}\right),\,
g=-\frac{1}{2E}\tanh\frac{Y_0E}{2}\cdot\Delta;\label{2}\end{equation}
$$E=\sqrt{\xi^2+\Delta\Delta^+},\;
\xi=\varepsilon+\frac{Y_4}{Y_0} $$
Here $Y_0=1/T$, $-Y_{4}/Y_0=\mu$, $T$ is  temperature,  $\mu$ is
chemical potential.

For a superfluid FL with the singlet-triplet (ST) pairing of fermions the
matrix order parameter has the form
\begin{equation}
\Delta({\mathbf p})=\Delta_{0}(\mathbf p)\sigma_2+
\vec{\Delta}(\mathbf p)\mathbf{\vec{\sigma}}\sigma_2,
 \label{3}\end{equation}
$\sigma_i$ being the Pauli matrices in spin space. The
quantities $\Delta_0$ and
 $\Delta_i$ in Eq.~\p{3} determine the singlet and triplet components of the
order parameter $\Delta$, respectively. In what follows, we shall assume that
the structure of $\Delta_0$ and $\Delta_i$ is such that
$\Delta_0(\mathbf p)=\Delta_0(p)$,
$\Delta_i(\mathbf p)=R_{ik}p_k^0\Delta(p)$, where $R_{ik}$ is a real rotation
matrix.
The wave function of a Cooper pair for singlet-triplet pairing
reads as
\begin{equation}
g({\mathbf p})=g_{0}(\mathbf p)\sigma_2+
\vec g(\mathbf p)\vec\sigma\sigma_2
\label{4}\end{equation}
To derive the self-consistent equations it is necessary to specify the energy
functional which we set in the form
\begin{equation}
E(f,g)=E_0(f)+E_{int}(g),\;
E_0(f)=2\sum_\mathbf p\varepsilon_0(\mathbf p)f_0(\mathbf p),
\quad\varepsilon_0(\mathbf
p)=\frac{{\mathbf p}^2}{2m}\label{5}\end{equation} $$
E_{int}(g)=\frac{1}{V}\sum_{\mathbf{pp'}}g_0^*(\mathbf p)L_s(\mathbf
p,\mathbf p') g_0(\mathbf p')+\frac{1}{V}\sum_{\mathbf{pp'}}\vec
g^{\, *}(\mathbf p)L_t(\mathbf p,\mathbf p') \vec g(\mathbf p')$$ Here $f_0$
is the coefficient of the unit matrix $\sigma_0$ in expansion of the
distribution function $f$ in the Pauli matrices; $L_s, L_t$ are the singlet
and triplet anomalous FL interaction amplitudes with the structure
 $L_s(\mathbf p,\mathbf p')=L_s(p,p'),L_t(\mathbf p,\mathbf p')=L_t(p,p')
\mathbf p^0\mathbf p'^{0}$.  The use of Eqs. \p{1},\p{2},\p{5} allows to
represent the self--consistent equations in the form \begin{equation} \Delta_ 0(p)=-\frac
{1}{ 4V}\sum\limits_{\mathbf q}^{} L_s(p, q )\left\{
\frac{\Delta_+(q)}{E_+(q)}\tanh\frac{Y_0E_+(q)}{2}+
\frac{\Delta_-(q)}{4E_-(q)}\tanh\frac{Y_0E_-(q)}{2}\right\},
\label{6}\end{equation}
$$\Delta
(p)=-\frac {1}{ 4V}\sum\limits_{\mathbf q}^{} \frac{L_t(p,q
)}{3}\left\{
\frac{\Delta_+(q)}{E_+(q)}\tanh\frac{Y_0E_+(q)}{2}-
\frac{\Delta_-(q)}{E_-(q)}\tanh\frac{Y_0E_-(q)}{2}\right\}$$
Here  $\Delta_\pm=\Delta_{0}\pm\Delta,\,E_\pm=\sqrt{\xi^2+\Delta_\pm^2}$.
  Eqs. \p{6} represent
themselves the system of the integral equations for determining the
singlet and triplet order parameters in the case of ST pairing of
quasiparticles in a superfluid FL.

\begin{figure}[t]
\begin{center}\leavevmode
\includegraphics[width=0.9\linewidth]{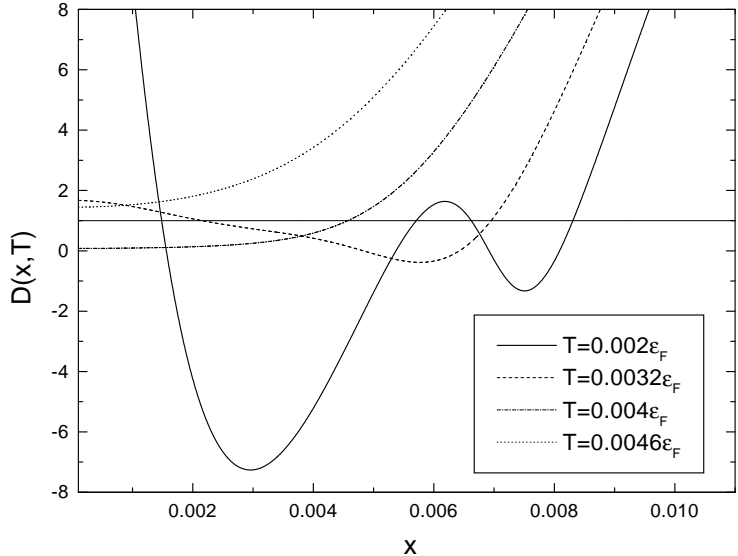}
\caption{Behaviour of the function
$D(x,T)$ at different temperatures.
The critical temperatures of various phase transitions in the model
case  with $g_s=0.25,g_t'=0.2$ and $\theta=0.01\varepsilon_F$
are as follows: $T_s=0.0045\varepsilon_F$, $T_t=0.0033\varepsilon_F$ and
$T_{st}=0.0028\varepsilon_F$.
}\label{fig0}\end{center}\end{figure}

Let us give an analysis of  Eqs. \p{6}, using the  model
representations of the BCS theory (the amplitudes $L_s$ and $L_t$ are not
equal to zero only in a narrow layer near the Fermi surface:
$|\xi|\leq\theta,\;\theta\ll\varepsilon_F$).  Then the quantities
$\Delta_0\equiv\Delta_0(p=p_F),$ $\Delta\equiv\Delta(p=p_F)$ can be found
from the relations
\begin{equation}\Delta_{0}=\frac{1}{2}x(1+d(x,T)),\;\Delta=
\frac{1}{2}x(1-d(x,T)) \label{7}\end{equation}
where $x$ is the solution of the equation \begin{equation}
d(x,T)\cdot d(x\cdot
d(x,T),T)\equiv D(x,T)=1,\label{8} \end{equation} $$
d(x,T)=\frac{4g_sg_t'\lambda(x,T)-g_s-g_t'}{g_t'-g_s},\;\;
\lambda(x,T)=\int_{-\theta}^\theta\frac{d\xi}{E}\tanh\frac{E}{2T},$$
$$E=\sqrt{\xi^2+x^2},\;
g_t'=\frac{g_t}{3},\; g_{s,t}=-
\frac{\nu_FL_{s,t}(p=p_F,q=p_F)}{4}$$
One--gap solutions are obtained  as solutions of the
equations $d(x,T)=1$ (singlet pairing) and $d(x,T)=-1$  (triplet pairing),
while two-gap solutions correspond to the case $d(x,T)\not=\pm1$
(singlet-triplet pairing). The critical temperatures of the
singlet and triplet superfluid  transitions are found from the equations
$d(0,T_{s})=1,\,d(0,T_{t})=-1$ respectively.
Assume for definiteness, that $g_s>g_t'$.  An analysis of the behaviour of
the function $D(x,T)$  shows  (see Fig.~\ref{fig0}) that equation $D(x,T)=1$
has no one-gap solutions at temperatures $T>T_s$, only one singlet
solution exists for $T_t<T<T_s$, while for $T_{st}<T<T_t$ the system is
characterized by two one-gap (singlet and triplet) solutions. Finally, at
$T<T_{st}$ we have two new ST solutions in addition to the previous
solutions.  For determining the critical temperature $T_{st}$ of transition
to the state with ST fermion pairing we have the equations \begin{equation}
d(x,T)=-1,\;\; xd'_x(x,T)=2 \label{9}\end{equation} The first
of these equations
indicates that ST solutions are continuously branched off the one-gap
triplet solution, while the second equation is the condition that the
derivative $D_x'(x,T)$ vanishes at the branching point.  Calculation of the
second derivative $D''_{xx}(x,T)$ in the critical point gives
$D''_{xx}(x_{st},T_{st})=0$, i.e., the mechanism of branching of ST solutions
is  the formation of inflection on the curve $z=D(x,T_{st})$ for $x=x_{st}$.

\begin{figure}[tbh]
\begin{center}\leavevmode
\includegraphics[width=0.9\linewidth]{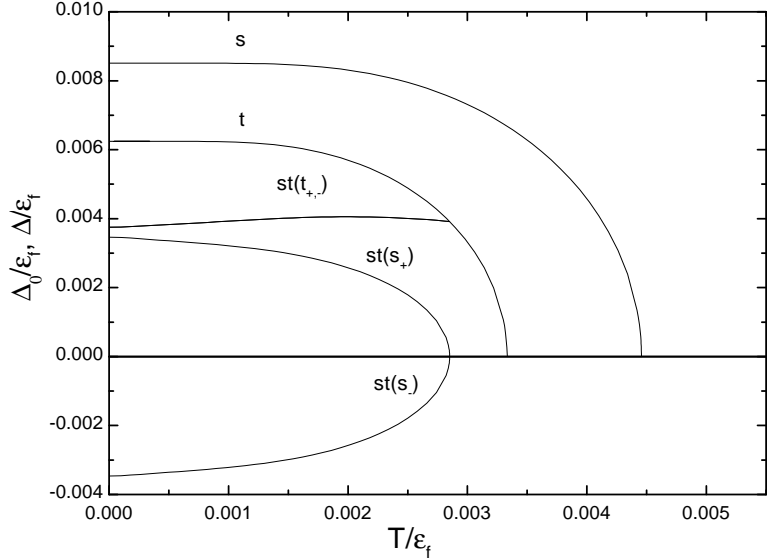} \caption{Order parameters
$\Delta_0$ and $\Delta$ vs temperature.  $st(s_+),st(s_-)$ and $st(t_{\pm})$
are notations for the dependences of the order parameters $\Delta_{0}$ and
$\Delta$ in two pair $(\pm\Delta_{0},\Delta)$ of ST solutions of
self-consistent equations.}\label{fig1}\end{center}\end{figure}

The numerical analysis of the equations for
determining $\Delta_0,\Delta$ in the
model case considered above  is given in Fig.~\ref{fig1}.   Note
that if the coupling constants satisfy the inequality $g_t'>g_s$ then  ST
solutions branch off the singlet one-gap solution.

\section{SYSTEMS OF FERMIONS OF TWO SORTS }
The problem of studying of phase transitions in nuclear matter has
encouraged a great interest (see Refs. \cite{ARS,BE} and
references therein).  However, earlier the normal-to-superfluid phase
  transition in nuclear matter has been studied only.  In principle, at
further lowering of  temperature a superfluid FL of nucleons can again become
unstable passing into a new superfluid state. Thus, we are speaking of the
superfluid-to-superfluid phase transitions in nuclear matter.
The basic formalism is laid out in more detail in Ref. \cite{AIP},
where the phase transitions to one-gap superfluid states of symmetrical
nuclear matter have been studied.  The main aspect now is analysis of the
phase transitions to
multi-gap superfluid states.
For simplicity we assume that the
energy functional is invariant under rotations in the configurational, spin
and isospin spaces.  Hence, superfluid phases (Cooper pairs) are
classified by specifying the total spin of the pair, $S=0,1$, the isospin
$T=0,1$, their projections $S_z$ and $T_z$ on the $z$ axis, and the orbital
angular momentum $L=0,1,2,...$    Each superfluid phase is described by its
own set of order parameters: at $S=0,\, T=0$ by the scalar order parameter
$\Delta_{00}$ (singlet-singlet pairing of nucleons), at $S=1,\,
T=0$ or $S=0,\,
T=1$ by the vector order parameter $\Delta_{k0}$ or
$\Delta_{0k}$, respectively
($k=1,2,3$) (triplet-singlet and singlet-triplet pairing of nucleons) and at
$S=1,\, T=1$ by the tensor order parameter $\Delta_{ik}$ ($i,k=1,2,3$)
(triplet-triplet pairing of nucleons).

Further we shall study the
unitary states of superfluid nuclear matter, for which the
product $\Delta\Delta^+$
is proportional to the product of the unit matrices in the spin and isospin
spaces, $\Delta\Delta^+\propto \sigma_0\tau_0$.  We consider as
a particular case
two-gap unitary states, for which the order parameter reads as
\begin{equation}
\Delta({\mathbf p})=\Delta_{30}(\mathbf p)\sigma_3\sigma_2\tau_2+
\Delta_{03}(\mathbf
p)\sigma_2\tau_3\tau_2 \label{10}\end{equation} In this case the
wave function of a Cooper
pair describes the superposition of states with the triplet-singlet and
singlet-triplet pairing of nucleons (TS-ST states) with the projections of
 total spin and isospin $S_z=T_z=0$.

To derive the self--consistent equations of a nucleon superfluid FL
it is
necessary to specify the energy functional, which we set in the form
 \begin{equation}
 E(f,g)=E_0(f)+E_{int}(g),\;\quad
{E}_0(f)=4\sum\limits_{ \mathbf p}^{}
\varepsilon_0(\mathbf p)f_{00}(\mathbf p),\; \varepsilon_0(\mathbf p)=
\frac{\mathbf p^{\,2}}{2m},
\label{12}
 \end{equation}
$${E}_{int}(g)=\frac{2}{V}\sum\limits_{ \mathbf p,\mathbf q}^{}
\{g_{30}^*( \mathbf p)V_1( \mathbf p, \mathbf q)g_{30}(\mathbf q) +
g_{03}^*(\mathbf p)V_2(\mathbf p,\mathbf q)g_{03}(\mathbf q)\}$$
Here $m$ is the effective nucleon mass, $f_{00}$ is the coefficient of the
product $\tau_0\sigma_0$ in expansion of the distribution function $f$ in the
Pauli matrices, $V_1,V_2$ are the anomalous FL interaction amplitudes, which
have the symmetry properties $V_i(-\mathbf p,\mathbf q)=
V_i(\mathbf p,-\mathbf q)=V_i(\mathbf p,\mathbf q), \,i=1,2$. The quantity
$m$ contains account of the normal FL effects and represents itself the mass
of a free nucleon, renormalized by the normal FL interaction \cite{AIP}.
The use of Eqs.
\p{1},\p{2},\p{12} allows to obtain the self--consistent equations in the
form of Eqs. \p{6}, where $\Delta_\pm=\Delta_{30}\pm\Delta_{03},\, E_\pm=
\sqrt{\xi^2+\Delta_\pm^2}$ and it is necessary to do the substitution
$L_s\rightarrow V_1,
L_t/3\rightarrow V_2$. In what follows, we shall choose the Skyrme forces
\cite{BGH,DFT} as the amplitudes of NN interaction
\begin{equation}
V_{1,2}(\mathbf p, \mathbf q)=t_0(1\pm x_0)+\frac{1}{6}t_3n^\alpha(1\pm x_3)
+\frac{1}{2\hbar^2}t_1(1\pm x_1)(\mathbf p^2 +\mathbf q^2),
\label{13}\end{equation}
where $n$ is the density of symmetrical nuclear matter, $t_i,x_i,\alpha$ are
some phenomenological parameters. Note that the amplitudes $V_1,V_2$ contain
no dependence from the
parameters $t_2,x_2$, because the amplitudes $V_1,V_2$ are the even functions
of the arguments $\mathbf p,\mathbf q$ (see details in Ref.
\cite{AIP}).  There are sets of parameters $t_i,x_i,\alpha$,
which differ
for various versions of the Skyrme forces (we shall use the Ska, SkM, SkM$^*$
and RATP potentials \cite{BGH} as well as the  SkP  potential \cite{DFT}).

\begin{figure}[bht]
\begin{center}\leavevmode
\includegraphics[width=0.9\linewidth]
{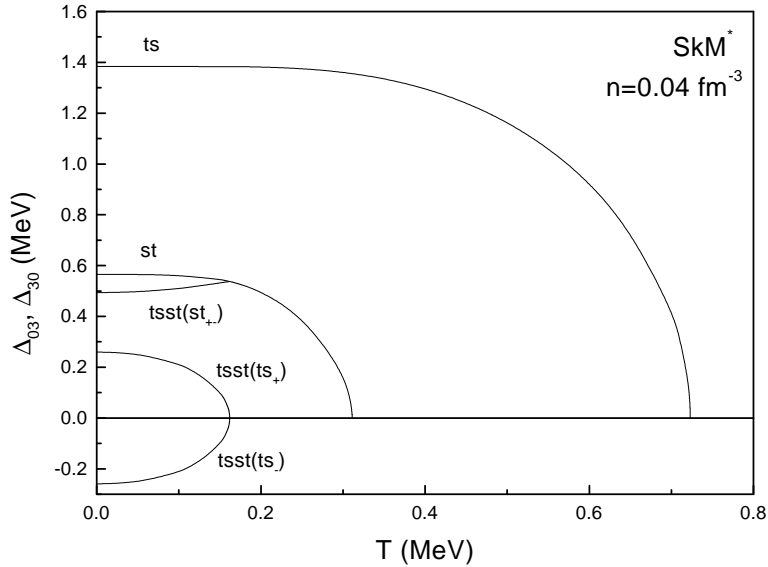}
\caption{Order parameters $\Delta_{30},\Delta_{03}$ vs temperature.
$tsst(ts_+),tsst(ts_-)$ and $tsst(st_{\pm})$ are notations for the
dependences of the  order parameters
in two pair $(\pm\Delta_{30},\Delta_{03})$ of TS-ST solutions of
 self-consistency equations.}
\label{fig2}\end{center}\end{figure}

   The temperature dependence of the order
parameters $\Delta_{30}\equiv\Delta_{30}(p=p_F),
\Delta_{03}\equiv\Delta_{03}(p=p_F)$ is
determined from Eq.~\p{7},\p{8} in which it is necessary to do the above
mentioned substitution of the amplitudes  as well as the substitution  of
the coupling constants $g_s\rightarrow g_1, g_t'\rightarrow g_2$. The results of numerical
integration for the potential SkM$^*$ at $n=0.4\rm{fm}^{-3}$
are given in
Fig.~\ref{fig2}.  It is seen that two-gap solutions branch off from one-gap
singlet-triplet solution. The results of calculations for the potentials SkM,
Ska have the form being analogous to the form in Fig.~\ref{fig2}; for the
potentials RATP and SkP the self-consistent equations have no two-gap
solutions.

\begin{figure}[thb]
\begin{center}\leavevmode
\includegraphics[width=0.9\linewidth]
{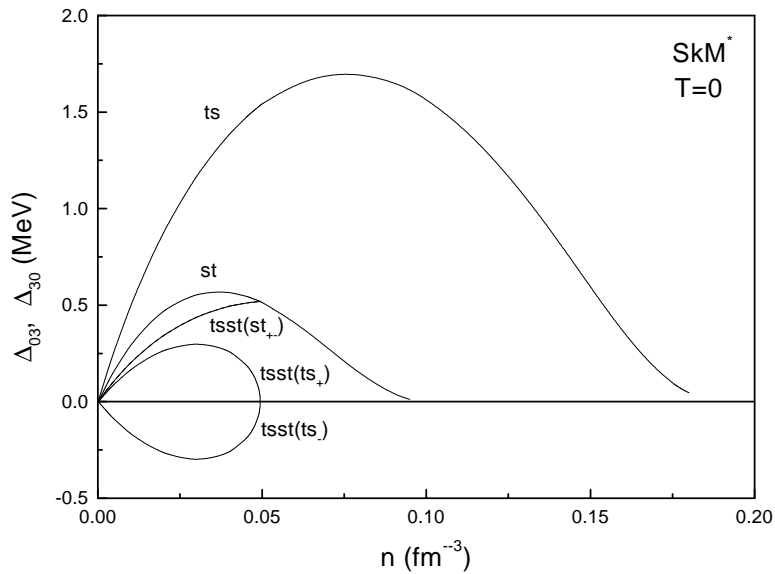}
\caption{Order parameters $\Delta_{30},\Delta_{03}$ vs density.
Notations are the same as in Fig.~\ref{fig2}.
}
\label{fig3}\end{center}\end{figure}

Note that since the coupling constants depend on density of nuclear matter,
there exists the possibility of phase transitions in density to multi-gap
states of superfluid nuclear matter as well. It is demonstrated on
Fig.~\ref{fig3} where the results of numerical determination of the order
parameters $\Delta_{30},\Delta_{03}$ as functions of density at $T=0$ are
presented. The results of calculations for the potentials SkM, Ska have
the same form whereas for the potentials RATP, SkP the self-consistent
equations have no two-gap solutions.
\section{CONCLUSION} Thus, we have pointed out the possibility
of phase transitions in superfluid FL, consisting of fermions of one
and two sorts, to two-gap superfluid states, corresponding to the
superposition of fermion pairings with different values of spin (and
isotopic spin).  The
self-consistent equations for these states differ essentially from the
equations of the BCS theory and contain the one-gap solutions  as
some particular cases. Phase transitions to multi-gap
superfluid states can take place either in temperature or density if the
coupling constants in various pairing channels depend on $n$. Note, that
 mixed pairing of quasiparticles (but not in spin space) was
considered in Ref.\cite{L} (superposition of $d_{x^2-y^2}$ and
$d_{xy}$
states in orbital space for HTSC) and for superconductors with
overlapping bands, first, in Ref.\cite{S} and in some
recent works (see, for example, Ref.\cite{A} and references therein).

\section*{ACKNOWLEDGMENTS}
This research is supported by BMBF.


\begin{thebibliography}{9}   \bibitem{AKP} A.I.  Akhiezer, V.V.
Krasil'nikov et al., {\it Phys.  Rep.}  {\bf 245}, 1
(1994).  \bibitem{AIP} A.I.  Akhiezer, A.A. Isayev, S.V.  Peletminsky et al.,
{\it Zh.  Eksp.  Teor.  Fiz.}  {\bf  112}, 3 (1997)  [{\it Sov. Phys. JETP}
{\bf 85}, 1 (1997) ].  \bibitem{ARS} Th.  Alm, G.  R\"opke, A.  Sedrakian
et al.,
{\it Nucl.  Phys.  A} {\bf 604}, 491 (1996).  \bibitem{BE} M. Baldo, \O.
Elgar{\o}y, L. Engvik et al., {\it Phys.  Rev.  C} {\bf 58}, 1921 (1998).
\bibitem{BGH} M.  Brack, C.  Guet and H.-B.  Hakansson,
{\it Phys.  Rep.}  {\bf
123}, 275 (1985).  \bibitem{DFT} J.  Dobaczewski, I.  Hamamoto et al.,
{\it Phys.  Rev.  Lett.} {\bf 72},  981 (1994).
\bibitem{L} R.B. Laughlin, {\it Phys.  Rev.  Lett.} {\bf 80},  5188 (1998).
\bibitem{S} H. Suhl, B.T. Matthias and L.R. Walker, {\it Phys. Rev. Lett.}
 {\bf 3},  552 (1959).
\bibitem{A} D.F. Agterberg, V. Barzykin and L. Gor'kov, cond-mat/9907331.

 \end{thebibliography}
\end{document}